\documentclass[twocolumn,showpacs,preprintnumbers,amsmath,amssymb,prd]{revtex4}

\usepackage{graphicx}
\usepackage{dcolumn}
\usepackage{bm}

\begin{document}

\preprint{}

\title{$U_{\mathrm{A}}(1)$ Anomaly in Hot and Dense QCD and the
Critical Surface}

\author{Jiunn-Wei Chen}
 \email{jwc@phys.ntu.edu.tw}
 \affiliation{
Department of Physics and Center for Theoretical Sciences, 
National Taiwan University, Taipei 10617, Taiwan
}%

\author{Kenji Fukushima}%
 \email{fuku@yukawa.kyoto-u.ac.jp}
 \affiliation{
Yukawa Institute for Theoretical Physics, Kyoto University, Kyoto 606-8502, Japan
}%

\author{Hiroaki Kohyama}
 \email{kohyama@phys.sinica.edu.tw}
 \affiliation{
Institute of Physics, Academia Sinica, Taipei, Taiwan 115, Republic of China
}
 \affiliation{
Physics Division, National Center for Theoretical Sciences,
Hsinchu, Taiwan 300, Republic of China
}%

\author{Kazuaki Ohnishi}%
 \email{kohnishi@phys.ntu.edu.tw}
 \affiliation{%
Department of Physics and Center for Theoretical Sciences, 
National Taiwan University, Taipei 10617, Taiwan
}%

\author{Udit Raha}%
 \email{udit@phys.ntu.edu.tw}
 \affiliation{%
Department of Physics and Center for Theoretical Sciences, 
National Taiwan University, Taipei 10617, Taiwan
}%

\date{\today}

\begin{abstract}
We discuss the chiral phase transition in hot and dense QCD with three light
flavors. Inspired by the well-known fact that the $U_{\mathrm{A}}(1)$\
anomaly could induce first order phase transitions, we study the effect of
the possible restoration of the $U_{\mathrm{A}}(1)$ symmetry at finite
density. In particular, we explore the link between the $U_{\mathrm{A}}(1)$
restoration and the recent lattice QCD results of de Forcrand and Philipsen,
in which the first order phase transition region near zero chemical
potential ($\mu $) shrinks in the quark mass and $\mu $ space when $\mu $ is
increased. Starting from the Ginzburg-Landau theory for general discussions,
we then use the Nambu--Jona-Lasinio model for quantitative studies. With the
partial $U_{\mathrm{A}}(1)$\ restoration modeled by the density dependent 't
Hooft interaction, we fit the shrinking of the first order region found in de
Forcrand and Philipsen's lattice calculation at low $\mu $. At higher $\mu $%
, the first order region might shrink or expand, depending on the scenarios.
This raises the possibility that despite the shrinking of the first order
region at lower $\mu $, the QCD critical end point might still exist due to
the expansion at higher $\mu $. In this case, very high precision lattice
data will be needed to detect the recently observed back-bending of the 
critical surface with
the currently available analytic continuation or Taylor expansion
approaches. Lattice computations could, however, test whether the $U_{%
\mathrm{A}}(1)$ restoration is responsible for the shrinking of the first order
region by computing the $\eta ^{\prime }$ mass or the topological
susceptibility at small $\mu $.
\end{abstract}

\maketitle


\section{\label{sec:intro}Introduction}

The $U_{\mathrm{A}}(1)$ anomaly is an interesting phenomenon of Quantum
Chromodynamics (QCD). In the chiral limit with three massless quark flavors,
QCD has the chiral $SU_{\mathrm{L}}(3)\otimes SU_{\mathrm{R}}(3)$ symmetry.
This symmetry is spontaneously broken in the QCD vacuum, giving rise to
eight massless Nambu-Goldstone bosons. The $U_{\mathrm{A}}(1)$ symmetry, on
the other hand, is broken not spontaneously but explicitly due to the
quantum anomaly. Thus, the $\eta ^{\prime }$ meson is not a Nambu-Goldstone
boson. It remains massive in the chiral limit. This is the so-called $U_{\mathrm{A}%
}(1) $ problem \cite{Weinberg:1975ui} and its consequence \cite{'tHooft:1986nc}%
. 't Hooft showed that instantons, which are topological configurations of
the classical gluon field, 
are related to the $U_{\mathrm{A}}(1)$ anomaly. He also constructed an
effective quark interaction which breaks the $U_{\mathrm{A}}(1)$ symmetry 
\cite{'tHooft:1976up} (see also \cite{Kobayashi:1970ji}). The coupling
constant of this (Kobayashi-Maskawa) 't Hooft interaction 
measures the effective strength of the $U_{\mathrm{A}}(1)$ anomaly relevant
to the hadron spectrum.

It is interesting to investigate the effective restoration of the $U_{%
\mathrm{A}}(1)$ symmetry at finite temperature ($T$) 
and/or quark chemical potential ($\mu $) 
induced by the decrease of instantons \cite{Pisarski:1983ms,Shuryak:1993ee},
even though the triangle anomaly is independent of the infrared scale $T$ 
\cite{Itoyama:1982up,Hsu:2000by}. Consequences of the $U_{\mathrm{A}}(1)$
restoration have been investigated if the restoration is associated with the
chiral transition \cite{SchaffnerBielich:1999uj,Kapusta:1995ww}. In Ref.\ 
\cite{Shuryak:1993ee}, based on the instanton liquid model, it was 
speculated that drastic $U_{\mathrm{A}}(1)$ restoration at the chiral
transition 
could be expected because the chiral transition is caused by the
rearrangement of the instanton configurations in that model.

An analysis using the Nambu--Jona-Lasinio (NJL) model was given in Ref.\ 
\cite{Fukushima:2001hr}, where the topological susceptibility $\chi_{t} $, a
correlator of topological charges, was considered. In the large $N_{\mathrm{c%
}}$ (number of colors) limit, $\chi_{t} $ is related to the $\eta ^{\prime }$
mass through the Witten-Veneziano mass formula \cite{Witten:1979vv}, $2N_{%
\mathrm{f}}\chi_{t} /f_{\pi }^{2}=m_{\eta }^{2}+m_{\eta ^{\prime
}}^{2}-2m_{K}^{2}$, so it can be used to probe the $U_{\mathrm{A}}(1)$
anomaly. The NJL model calculation \cite{Fukushima:2001hr} reproduced the
lattice data \cite{Alles:1996nm} above the critical temperature up to 1.5
times the chiral phase transition temperature with temperature independent
't Hooft coupling constant. This implies that, at least in the NJL model,
the effective $U_{\mathrm{A}}(1)$ restoration does not necessarily take
place near the chiral transition.

At finite density, the medium response of the $U_{\mathrm{A}}(1)$ anomaly is
even less understood, largely because of the lack of lattice data due to the
fermion sign problem \cite{Alles:2006ea}. Reliable analyses are available only at
asymptotically high density where QCD is perturbative. In Refs.\ \cite%
{Manuel:2000wm,Son:2001jm,Schafer:2002ty}, the $\eta ^{\prime }$ mass in the
Color-Flavor Locked (CFL) phase was calculated and shown to decrease as the
density increases. This corresponds to the realization of effective $\mathrm{%
U}_{\mathrm{A}}(1)$ restoration. 
There are efforts to test this restoration hypothesis by probing the
decrease in the $\eta $\ and $\eta ^{\prime }$\ masses \cite%
{Pisarski:1983ms,Kikuchi:1987jr,Hashimoto:1987bs,Alkofer:1989rr,Kunihiro:1989my, Costa:2002gk}
in medium. It is claimed that the decrease can be studied experimentally
through observing the possible formation of $\eta $- and $\eta ^{\prime }$%
-mesic nuclei \cite%
{Haider:1986sa,Tsushima:1998qw,Hayano:1998sy,Inoue:2002xw,Nagahiro:2004qz,Bass:2005hn}%
. 

It is interesting to speculate how the 
anomalous violation and restoration of $U_{\mathrm{A}}(1)$ symmetry could
change the QCD phase diagram (see Ref.\ \cite{Hatsuda:2006ps} for example).
In this work, we will not discuss the possible changes of phases and phase
boundaries in the intermediate density. Instead, we will focus on the issue
regarding the existence of the QCD\ critical end point (CEP). In the
conventional three flavor picture
\cite{Asakawa:1989bq,Barducci:1989wi,Stephanov:1998dy}, 
there is a first order boundary of the chiral phase transition separating
the hadronic and quark phases starting from a point with zero $T$ but
non-zero $\mu $ to a point ($\mu _{\mathrm{c}},T_{\mathrm{c}}$), then the
QCD phase transition becomes a crossover at lower $\mu $. This end point of
the first order phase transition, ($\mu _{\mathrm{c}},T_{\mathrm{c}}$), is
called the CEP. This CEP has an second order phase transition despite the
finite quark masses. The search of the QCD CEP is a 
priority in the next phase of the RHIC running.

However, this conventional picture of the QCD phase diagram and the
existence of the CEP are challenged by the recent lattice QCD results of
de Forcrand and Philipsen \cite{deForcrand:2006pv}. The assertion in Ref.\ 
\cite{deForcrand:2006pv} can be explained by the so-called Columbia plot,
where each point in the parameter space of quark masses and $\mu $\ is
marked by its order of phase transition when $T$\ is increased. We will
study QCD with three light flavors, $u,d$\ and $s$, with isospin symmetry $%
m_{\mathrm{u}}=m_{\mathrm{d}}\equiv m_{\mathrm{ud}}$. When $\mu =0$, the
chiral limit point ($m_{\mathrm{ud}}=m_{\mathrm{s}}=0$) has a first order
phase transition when $T$\ is increased due to symmetry reasons \cite%
{Pisarski:1983ms}. Around the chiral limit, there exists a finite area
such that each point in this area represents a 
first order phase transition. The boundary of this
area is marked by a critical curve. Each point on this critical curve is of
a second order phase transition. Beyond the critical curve, the theory has a
crossover until all the quarks become heavy such that the theory is close to
a pure Yang-Mills theory 
and phase transitions can then take place again. In the following, we will
just focus on the light quark region.

At $\mu =0$, the physical point of quark masses is located in the crossover
region, meaning that QCD has a crossover at $\mu =0$. If the CEP exists at
finite $\mu $, then 
the physical point should enter the first order phase transition region at
finite $\mu $. However, in Ref. \cite{deForcrand:2006pv}, it was found that
near $\mu =0$, the 
region of the first order phase transition shrinks as $\mu $\ increases.
Note that this computation is not directly carried out 
with finite $\mu $ QCD due to the fermion sign problem. Instead, 
methods with analytical continuation from imaginary $\mu $ and with
derivatives computed at $\mu =0$ are used. Thus, essentially it is the
curvature 
at $\mu =0$\ that was computed. Although that result of Ref.\ \cite%
{deForcrand:2006pv} disfavors the existence of the CEP, to give a definite
answer to 
whether the CEP exists or not, computations at higher $\mu $\ are necessary
(note that other lattice results using different approaches are consistent
with the existence of the CEP \cite{FK,Ejiri,Gavai,Ejiri:2006ft}).

Given that lattice computations at higher $\mu $\ are still challenging, it
is our hope that model calculations might shed light on this problem.

Inspired by the 
fact that the $U_{\mathrm{A}}(1)$\ anomaly could induce the first order
transition in massless three-flavor QCD, we study the effect of the possible
restoration of the $U_{\mathrm{A}}(1)$\ symmetry at finite density. Starting
from general discussions using the Ginzburg-Landau theory, we then use the 
NJL model for quantitative studies. With the partial $U_{\mathrm{A}}(1)$\
restoration modeled by the density dependent 't Hooft interaction, we fit
the negative curvature of 
the critical surface at $\mu =0$\ obtained in Ref.\ \cite{deForcrand:2006pv}%
. Finally, we discuss the behavior at higher $\mu $ within this model.

\section{\label{sec:GL}Ginzburg-Landau theory analysis}

In this section, we use the Ginzburg-Landau (GL) theory to demonstrate the
special role of the $U_{\mathrm{A}}(1)$\ anomaly in chiral\ phase
transitions, following the framework of Refs.\ \cite%
{Pisarski:1983ms,Hatsuda:2006ps,Gavin:1993yk,Hatta:2002sj,Yagi:2005yb}.

The effective potential $\Omega _{GL}$\ for the order parameter (chiral
condensate) field $\Phi _{ij}\equiv \left\langle q_{i}(1-\gamma _{5})%
\overline{q}_{j}\right\rangle $\ in QCD with three flavors ($N_{\mathrm{f}%
}=3 $) is parametrized by a set of operators satisfying the $SU_{\mathrm{L}%
}(3)\otimes SU_{\mathrm{R}}(3)\otimes U_{\mathrm{V}}(1)\otimes Z(N_{\mathrm{f%
}})$ symmetry, where $Z(N_{\mathrm{f}})$\ is the remaining symmetry of $U_{%
\mathrm{A}}(1)$\ after it is broken by quantum anomaly. Here, we will use
the mean field approximation and neglect the space time dependence of $\Phi $%
. This approximation neglects 
soft-mode fluctuations, which are large near second order phase transitions,
but it is still useful 
to explore the phase structure. Also, we will Taylor expand $\Omega _{GL}$
in analytic functions of the order parameter. 
The following analysis 
helps us to understand our NJL result in later discussions. We have 
\begin{eqnarray}
\Omega _{GL} &=&\frac{a_{0}}{2}\mathrm{Tr}\Phi ^{\dagger }\Phi +\frac{b_{1}}{%
4!}\left( \mathrm{Tr}\Phi ^{\dagger }\Phi \right) ^{2}+\frac{b_{2}}{4!}%
\mathrm{Tr}\left( \Phi ^{\dagger }\Phi \right) ^{2}  \nonumber \\
&&-\frac{c_{0}}{2}\left( \mathrm{det}\Phi +\mathrm{det}\Phi ^{\dagger
}\right) -\frac{1}{2}\mathrm{Tr}h_{0}\left( \Phi +\Phi ^{\dagger }\right) ,
\label{GL1}
\end{eqnarray}%
where the last term breaks chiral symmetry explicitly with $h_{0}\propto 
\mathrm{diag}(m_{\mathrm{u}},m_{\mathrm{d}},m_{\mathrm{s}})$. Adding higher
dimensional operators does not change the analysis qualitatively. The
determinant term simulates the $U_{\mathrm{A}}(1)$\ anomaly and has the $%
Z(N_{\mathrm{f}})$\ symmetry. We have $c_{0}\geq 0$\ at $T=0$ to yield a
finite $\eta ^{\prime }$\ mass in the chiral limit.

For simplicity, we restrict ourselves to the flavor $SU(3)$\ symmetric case: 
$m_{\mathrm{u}}=m_{\mathrm{d}}=m_{\mathrm{s}}$ (that is, on the diagonal in
the Columbia plot). Then we have $\Phi =\mathrm{diag}(\sigma ,\sigma ,\sigma
)$. The GL functional is reduced to 
\begin{eqnarray}
\Omega _{GL} &=&\frac{3}{2}a_{0}\sigma ^{2}-c_{0}\sigma ^{3}+\left( \frac{9}{%
4!}b_{1}+\frac{3}{4!}b_{2}\right) \sigma ^{4}-3h_{0}\sigma  \nonumber \\
&\equiv &\frac{1}{2}a\sigma ^{2}-\frac{1}{3}c\sigma ^{3}+\frac{1}{4}b\sigma
^{4}-h\sigma \ ,  \label{GL2}
\end{eqnarray}%
where $b>0$\ such that the free energy is bounded from below.

\begin{description}
\item[(i)] Chiral limit ($h=0$)
\end{description}

We first consider the chiral limit case with $h=0$, where interesting
results can be obtained.

If $a>0$\ and $c\neq 0$\ at the phase transition temperature, then Eq.\ (\ref%
{GL2}) has two local minima and the phase transition is of first order. The
first order transition 
persists against 
an external field $h$\ 
until it is washed out at sufficiently large $h$. Thus, the first order
region has some 
finite extent from $m_{\mathrm{ud}}=m_{\mathrm{s}}=0$ in the Columbia plot.
Going back to 
$h=0$, the two minima are located at $\sigma _{1}=0$\ and $\sigma _{2}=(c+%
\sqrt{c^{2}-4ab})/2b$. The chiral condensate is 
\begin{equation}
\Delta \sigma \equiv \sigma _{2}-\sigma _{1}=\frac{c+\sqrt{c^{2}-4ab}}{2b}.
\label{gap}
\end{equation}%
Note 
that as $c\rightarrow 0$, if there is a phase transition, then $a\rightarrow
0$\ at the phase transition point and the phase transition becomes second
order. Here we have assumed as $c\rightarrow 0$, $b$\ is still positive and
that the higher order terms in $\sigma $\ can further be neglected.
Otherwise first order phase transition is still possible through the
inclusion of the higher order terms such as $\sigma ^{6}$. Only in this simple 
case, the absence of the $U_{\mathrm{A}}(1)$\ anomaly 
leads to the disappearance of the first order transition.

For simplicity let us assume an (unphysical) extreme case that the $\mu $
dependence lies only in the anomaly term $c$ and $c\rightarrow 0$ at $\mu
=\mu _{\mathrm{res}}$. 
Then as we go from $\mu=0$ to higher $\mu$ on the $\mu$ axis, the
strength of the first order transition gets weakened. At $\mu=\mu _{\mathrm{res}}$,
the first order transition disappears completely, and above
$\mu _{\mathrm{res}}$, the transition turns into second order.
However, if fluctuations are taken into account,
the above mean-field picture is modified. The second order transition above $\mu
_{\mathrm{res}}$ has the $SU_{\mathrm{L}}(3)\otimes SU_{\mathrm{R}%
}(3)\otimes U_{\mathrm{A}}(1)$ symmetry. The renormalization group analysis
tells us that there is no infrared stable fixed point for this universality
class \cite{Pisarski:1983ms}. This means that the fluctuations wash out the
critical point and make it a fluctuation induced first order transition.
Thus, at $m_{\mathrm{ud}}=m_{\mathrm{s}}=0$, the transition remains first
order on the $\mu $\ axis even if the $U_{\mathrm{A}}(1)$ symmetry is restored.
This implies that as we go to $\mu _{\mathrm{res}}$, the first order region around
$m_{\mathrm{ud}}=m_{\mathrm{s}}=0$ would ever shrink, but the critical surface
can never touch the $\mu $ axis but only approaches it asymptotically.

The above discussion gives a clear picture about the eventual shrinking of
the first order region if $U_{\mathrm{A}}(1)$\ is completely restored above a
certain $\mu $ (if terms higher order than Eq.\ (\ref{GL2}) can be
neglected). For partial restoration with the GL coefficients depending on $%
\mu $, the eventual shrinking does not have to happen. This can be seen from 
$\Delta \sigma $\ in Eq.\ (\ref{gap}). If $c$\ decreases, $\Delta \sigma $\
can still increase if the parameter $b$ (or $a$) decreases sufficiently
fast. 
This means that even if the $U_{\mathrm{A}}(1)$\ symmetry is partially
restored when $\mu $\ increases, the first order region 
may still expand. In fact, this is confirmed in the NJL calculation in the
next section. 
In contrast, even without the $U_{\mathrm{A}}(1)$\ symmetry restoration, the
critical surface can still shrink if $b$ or $a$ increases due to some
mechanism. This is similar to what happens if a repulsive interaction
between\ vector currents is added to the NJL model, as demonstrated in Ref.\ 
\cite{Fukushima:2008is}. Thus, the $U_{\mathrm{A}}(1)$\ restoration is just one
possible explanation to the shrinkage of the first order region at small $\mu $%
.



\begin{description}
\item[(ii)] Finite current quark mass ($h\neq 0$)
\end{description}

The different scenarios mentioned above can also be seen in the following
analysis involving finite quark masses. At the critical end point, the GL
functional takes the form of 
\begin{eqnarray}
\Omega_{GL} &=&\frac{1}{4}b\left( \sigma -\sigma _{0}\right) ^{4}+d 
\nonumber \\
&=&\frac{1}{4}b\left( \sigma ^{4}-4\sigma _{0}\sigma ^{3}+6\sigma
_{0}^{2}\sigma ^{2}-4\sigma _{0}^{3}\sigma +\sigma _{0}^{4}\right) +d.
\end{eqnarray}%
Comparing this with Eq.\ (\ref{GL2}), we have 
\begin{equation}
b\sigma _{0}=\frac{1}{3}c,\hspace{3mm}\frac{3}{2}b\sigma _{0}^{2}=\frac{1}{2}%
a,\hspace{3mm}b\sigma _{0}^{3}=h,\hspace{3mm}\frac{1}{4}b\sigma _{0}^{4}+d=0.
\label{GL3}
\end{equation}%
These are the conditions to determine the critical surface. However, there
are six parameters--- $a$, $b$, $c$, $h$, $\sigma _{0}$ and $d$, while there
are only four relations among them in Eq.\ (\ref{GL3}). Thus, as mentioned
above, one cannot identify what causes the shrinking of the first order region
without extra inputs. 

\section{\label{sec:NJL}NJL model Analysis}

In this section, we give a quantitative analysis using the NJL model \cite%
{Klevansky:1992qe,Hatsuda:1994pi,Buballa:2003qv} with the partial $U_{%
\mathrm{A}}(1)$\ restoration modeled by the density dependent 't Hooft
interaction. We will fit the curvature of the shrinkage of the first order
region at $\mu =0$\ to the lattice QCD result of Ref. \cite%
{deForcrand:2006pv}. We then discuss the behaviors at higher $\mu $.

\subsection{Model Setting}

The NJL Lagrangian is given by 
\begin{eqnarray}
\mathcal{L} &=&\mathcal{L}_{0}+\mathcal{L}_{4}+\mathcal{L}_{6}, \\
\mathcal{L}_{0} &=&\bar{q}\left( \mathrm{i}\gamma \cdot \partial -\hat{m}\right) q, \\
\mathcal{L}_{4} &=&\frac{g_{\mathrm{S}}}{2}\sum_{a=0}^{8}\left[ \left( \bar{q}\lambda _{a}q\right) ^{2}+\left( \bar{q}\mathrm{i}\gamma _{5}\lambda_{a}q\right) ^{2}\right] , \\
\mathcal{L}_{6} &=&g_{\mathrm{D}}^{\ast }\left[ \mathrm{det}\,\bar{q}_{i}(1-\gamma _{5})q_{j}+\mathrm{h.c.}\right] .
\end{eqnarray}%
The kinetic term $\mathcal{L}_{0}$ includes the current quark mass matrix $%
\hat{m}$\ which breaks chiral symmetry explicitly. The sum of the two four
Fermion contact interaction terms in $\mathcal{L}_{4}$ is chirally
symmetric. It gives the attractive interaction responsible for the
spontaneous chiral symmetry breaking when its strength exceeds a certain
critical value. In this minimal model, the other types of four Fermion
contact interactions of the same mass dimension are not
included because they are heavier excitations below the phase transition. 
The six Fermion determinant term is the 't Hooft interaction which breaks
the $U_{\mathrm{A}}(1)$ symmetry.

The theory has five parameters: $m_{\mathrm{ud}}$, $m_{\mathrm{s}}$, $g_{%
\mathrm{S}}$, $g_{\mathrm{D}}^{\ast }$, and the cut-off $\Lambda $. They 
fit the following physical quantities in the vacuum: $m_{\pi }=138$ MeV, $%
f_{\pi }=93$ MeV, $m_{K}=495.7$ MeV, and $m_{\eta ^{\prime }}=957.5$ MeV
with $m_{\mathrm{ud}}=5.5$ MeV fixed. The resulting values are $m_{\mathrm{s}%
}=135.7$ MeV, $g_{\mathrm{S}}=3.67\Lambda ^{-2}$, and $\Lambda =631.4$ MeV 
\cite{Hatsuda:1994pi}. Note that in this paper, the 't Hooft coupling
constant is determined by fitting the $\eta ^{\prime }$ mass,
$g_{\mathrm{D}}^{\ast }=-9.29\Lambda ^{-5}\equiv g_{\mathrm{D}}$,
where we have defined $g_{\mathrm{D}}$ as the value of 
$g_{\mathrm{D}}^{\ast }$ at the vacuum.
Alternatively, one can choose to fit it to the mass boundary of the critical
surface using lattice results.

We will follow the standard procedure to calculate the effective potential
of the NJL model in the mean-field approximation 
\cite{Hatsuda:1994pi,Klevansky:1992qe,Buballa:2003qv,Asakawa:1989bq}. From the
effective potential one can construct the critical surface. The technical
details will be omitted here.
\begin{figure}[tbh]
\begin{center}
\includegraphics[height=2.5in,keepaspectratio,angle=0]{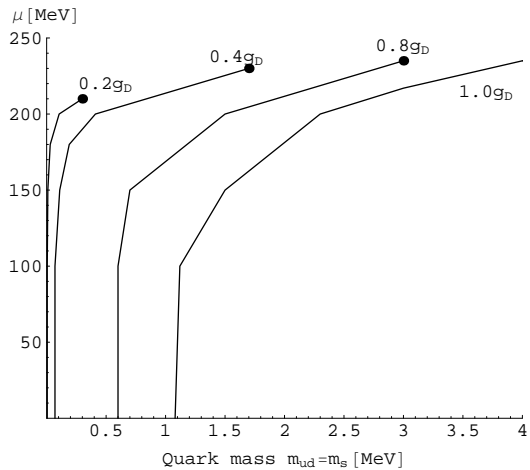}
\end{center}
\caption{The critical curves of constant $g_{\mathrm{D}}^{*}$ with 
$g_{\mathrm{D}}^{*}=g_{\mathrm{D}},0.8g_{\mathrm{D}},0.4g_{\mathrm{D}},$
and $0.2g_{\mathrm{D}}$, respectively. If $g_{\mathrm{D}}^{*}$ is a
monotonically decreasing function of $\mu $, then the critical curve
will bend to the left (corresponding to shrinking of the first order region)
at small $\mu $, as the lattice result of \cite{deForcrand:2006pv} has 
shown. At higher $\mu $, whether the
curve can keep bending to the left or eventually bend to the right
(corresponding to shrinking or eventual expansion of the first order region),
depends on the functional form of $g_{\mathrm{D}}^{* }(\mu )$.}
\label{const_gD}
\end{figure}

To apply the NJL model to finite $T$ and $\mu $, we will include the $\mu $
dependence in $g_{\mathrm{D}}^{\ast }$, $g_{\mathrm{D}}^{\ast}=g_{\mathrm{D}%
}^{\ast }(\mu )$, while the other parameters $m_{\mathrm{ud}}$, $m_{\mathrm{s%
}}$, $g_{\mathrm{S}}$, and $\Lambda $ remains independent of $T$ and $\mu $.
The $T$ dependence of $g_{\mathrm{D}}^{\ast }$ is small in this model and is
thus neglected. Note that in Ref.\ \cite{Fukushima:2001hr}, the $T$ independent
$g_{\mathrm{D}}^{\ast }$ was found to be able to reproduce the $T$ dependence
of the topological susceptibility. 
Without knowing the functional form of $g_{\mathrm{D}}^{\ast }(\mu )$, we
plot the critical curves in the ($\mu ,m=m_{\mathrm{ud}}=m_{\mathrm{s}}$)
space with different constant $g_{\mathrm{D}}^{\ast }$'s in Fig.\ \ref%
{const_gD} (the Columbia plot with $g_{\mathrm{D}}^{\ast }(\mu )=g_{\mathrm{%
D}}$ is shown in Fig. 3 of Ref.\ \cite{Fukushima:2008wg}.) Each curve ends
with a point beyond which first order phase transition ceases to exist, but 
crossover transition can still happen. This plot shows that the first order
regions expand as $\mu $ increases and larger the value of $g_{\mathrm{D}}^{\ast}$, larger the first order region. One immediately realizes, if $g_{%
\mathrm{D}}^{\ast }$ is a monotonically decreasing function of $\mu $, then
the critical curve will bend to the left (corresponding to the shrinking of
the first order region) at small $\mu $, as the lattice result of \cite{deForcrand:2006pv} has shown.
But, at higher $\mu $, whether the 
critical surface could keep shrinking or rather expands, depends on the
functional form of $g_{\mathrm{D}}^{\ast }(\mu )$, i.e., how
fast the $U_{\mathrm{A}}(1)$ symmetry gets restored.

As a special example, we will use the ansatz,%
\begin{equation}
g_{\mathrm{D}}^{\ast }(\mu )=g_{\mathrm{D}}^{\ast }(0)\,\mathrm{e}^{-\mu
^{2}/\mu _{0}^{2}}\ ,  \label{anzatz}
\end{equation}%
with one free parameter $\mu _{0}$ to parametrize the $U_{\mathrm{A}}(1)$
symmetry restoration. This form is motivated by the Gaussian suppression of
the instanton density due to Debye screening \cite%
{AragaodeCarvalho:1980de,Baluni:1980db,Shuryak:1982hk,Abrikosov:1983rh,Schafer:1996wv}%
.

\subsection{Lattice Data}

Before going to the numerical results of the NJL model, we discuss the
lattice data of the critical surface \cite{deForcrand:2006pv}. 
The critical curve, which is the intersect of the $SU(3)$ 
symmetric plane, $m=m_{\mathrm{ud}}=m_{\mathrm{s}}$, to the critical surface
can be Taylor expanded near $\mu =0$, in even powers due to CP symmetry: 
\begin{equation}
\frac{m_{\mathrm{c}}(\mu )}{m_{\mathrm{c}}(0)}=1-3.3(3)\left( \frac{\mu }{%
\pi T_{\mathrm{c}}}\right) ^{2}-47(20)\left( \frac{\mu }{\pi T_{\mathrm{c}}}%
\right) ^{4}-\cdots  \label{LQCD}
\end{equation}%
It is not clear what the radius of convergence of the above expansion is. By
requiring the third term to be smaller than the second term (which are both $%
\sim 23\%$ of the first term), we obtain $\mu \lesssim 90$-$100$\ MeV.

The negative signs for the $\mu ^{2}$\ and $\mu ^{4}$\ terms suggest the
shrinking of the first order region at small $\mu $. The sign of the $\mu ^{6}$%
\ term was also asserted to be negative, making the shrinking even more
serious, although the value of the coefficient has not yet been determined 
\cite{deForcrand:2006pv}. However, as we discussed above, we do not expect
the critical surface to touch the $\mu $\ axis at zero quark masses. Thus,
there must be terms at higher order in the $\mu $\ expansion with positive
prefactors which change 
the critical surface from the charecteristic front bending to back bending
at some density.

\subsection{Numerical Results in the NJL Model}

Now let us discuss the numerical results of the NJL model. In Fig.\ \ref%
{2D_eta_gD}, we show the result of the critical curve in the ($\mu $, $m=m_{%
\mathrm{ud}}=m_{\mathrm{s}}$) plane for various values of the free parameter 
$\mu _{0}$ \cite{Schaefer:2008hk}.
\begin{figure}[tbh]
\begin{center}
\includegraphics[height=3.0in,keepaspectratio,angle=0]{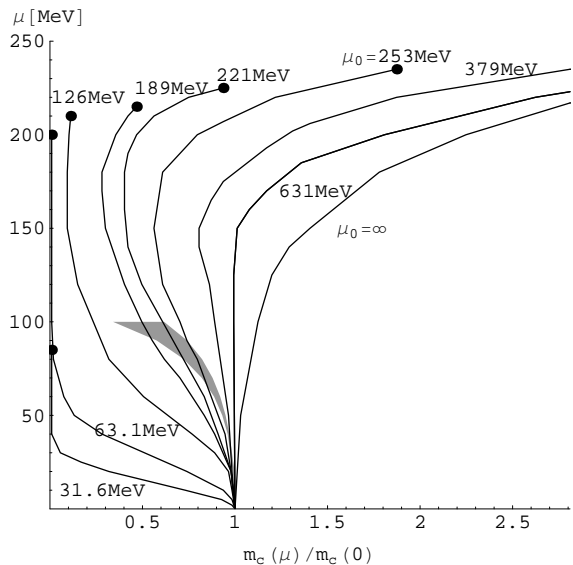}
\end{center}
\caption{The critical curves in the ($\protect\mu $,$m=m_{\mathrm{ud}}=m_{%
\mathrm{s}}$) plane with $g_{\mathrm{D}}^{\ast }\left( 0\right) =g_{\mathrm{D%
}}$ for several values of $\protect\mu _{0}$. The shaded region represents
the lattice result of Eq.\ (\protect\ref{LQCD}).
The lattice value of $T_{\mathrm{c}}$ depends on $\mu$. At $\mu=0$,
$T_{\mathrm{c}}=135$ MeV which gives a typical value for $T_{\mathrm{c}}$.
}
\label{2D_eta_gD}
\end{figure}
The shaded 
region represents the lattice result of Eq.\ (\ref{LQCD}). The band shows
the errors 
in the expansion coefficients but not from the higher order terms
or the systematic errors.

Several curves are plotted in the figure. As we go from the left-hand curve
to the right-hand one, $\mu _{0}$ increases from $31.6$ MeV to infinity. The
infinite $\mu _{0}$ corresponds to the density independent $g_{\mathrm{D}%
}^{\ast }$. The critical curves terminate at the critical chemical
potentials at $T=0$, above which first order phase transition ceases to
exist, but 
crossover transition can still happen. We see that the values of $\mu _{0}$
can be divided into three regions according to the qualitative behaviors of
the corresponding critical curves.

(i) The small $\mu _{0}$ region: In this region, the suppression of the $U_{%
\mathrm{A}}(1)$ anomaly is so strong that the first order region shrinks quite
rapidly. The surface keeps shrinking and approaches the $\mu $ axis
asymptotically.

(ii) The intermediate $\mu _{0}$ region: At low density, the first order
region shrinks and then expands again at high density. This back-bending
behavior is due to the density effect. In general, stronger first order
transition is favored at higher density, as discussed in detail in Ref.\ 
\cite{Fukushima:2008is}. In the language of the GL theory, this means that
the higher order terms enter the thermodynamic potential. 
Thus at finite density, there is a competition between the $U_{\mathrm{A}%
}(1) $ restoration which makes $c$ smaller and the density effect which 
generates higher order terms. In this intermediate $\mu _{0}$ region, the
density effect overcomes the $U_{\mathrm{A}}(1)$ restoration at high
density, resulting in the back-bending structure.

(iii) The large $\mu _{0}$ region: When $\mu _{0}$ is large, the $U_{\mathrm{%
A}}(1)$ restoration is weak. Thus, the first order region does not 
shrink and instead
expands monotonically. This corresponds to the conventional scenario with
constant $g_{\mathrm{D}}^{\ast }$ in which the CEP could exist.

Now let us compare the curves with the lattice data. We see that the curves
with $\mu _{0}=253\sim 379\ $MeV describes the lattice data well for $\mu
\lesssim 70$ MeV.
It is notable that this value of $\mu_0$ is close to a rough estimate by
$\mu_0\sim 1/(\sqrt{N_{\rm f}}\rho_0)\sim 380\,$MeV, where $N_{\rm f}=3$ and
$\rho_0\simeq0.3\ $fm is the typical instanton size. For 70 MeV $\lesssim \mu \lesssim 100$ MeV,
$\mu_{0}=189\sim 253\ $MeV describes the lattice data better.

\begin{figure}[tbh]
\begin{center}
\textbf{\includegraphics[height=3.0in,keepaspectratio,angle=0]{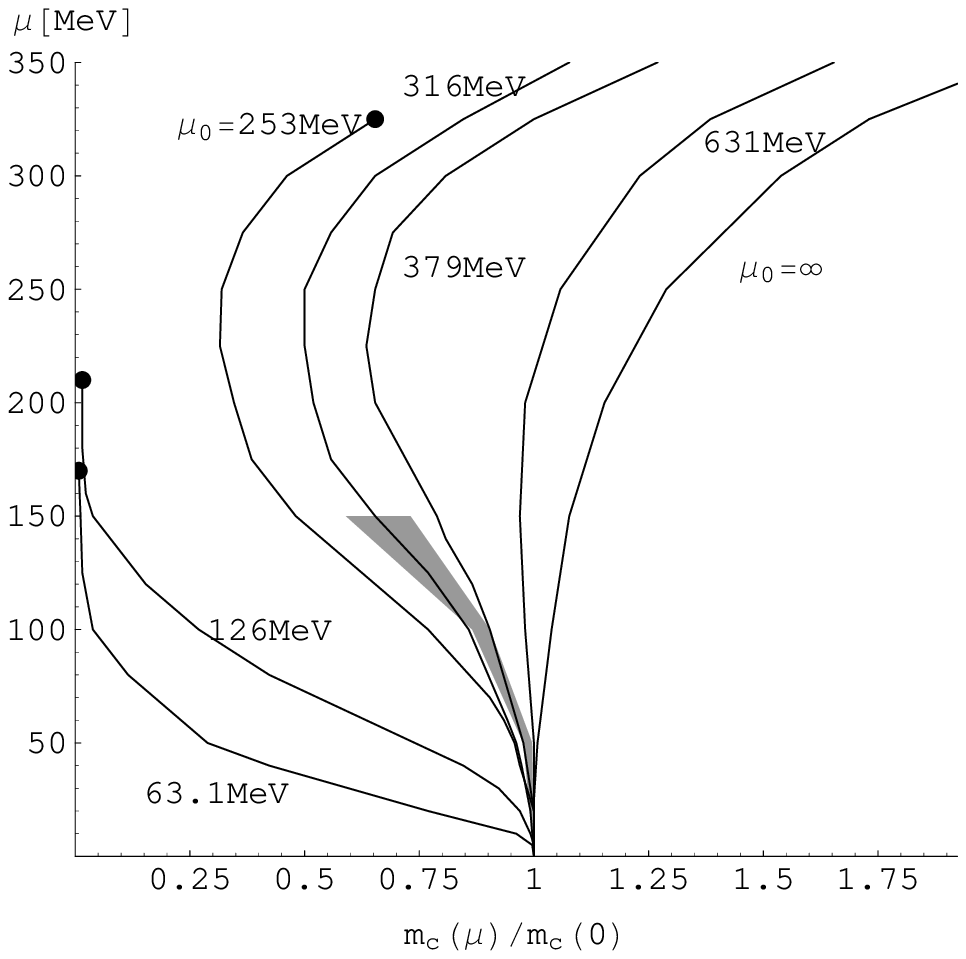} }
\end{center}
\caption{Same as Fig.\ \protect\ref{2D_eta_gD} but with $g_{\mathrm{D}%
}^{\ast }\left( 0\right) =3.5g_{\mathrm{D}}$. }
\label{2D_35gD}
\end{figure}

In our calculation $m_{\mathrm{c}}(0)\sim 1.1$ MeV is obtained while the
lattice data gives $m_{\mathrm{c}}(0)\sim 14$ MeV (extracted from Fig. 9 of
the first paper in Ref. \cite{deForcrand:2006pv}). One can argue that since $%
m_{\eta ^{\prime }}$ is heavier than the cutoff $\Lambda $ of the theory,
it should not be used to fit the parameters of the theory. Instead, one can
use the $m_{\mathrm{c}}(0)$ computed in lattice QCD. To explore the effects
of this new set of parameters, we use $g_{\mathrm{D}}^{\ast }\left( 0\right)
=3.5g_{\mathrm{D}}$ and keep the other parameters the same. This parameter
set gives $m_{\mathrm{c}}(0)\sim 26$ MeV, not quite the same as the lattice $%
m_{\mathrm{c}}(0)$, but rather the Columbia plot at $\mu =0$ is very similar to
that of \cite{deForcrand:2006pv}. In Fig.\ \ref{2D_35gD}, we plot the
corresponding critical curves. The qualitative features are the same as
those in Fig. \ref{2D_eta_gD}: There is a range of $\mu _{0}$ ($253\sim 379\ 
$MeV) where the curvature from lattice QCD is reproduced. In the 
range below $\mu _{0}=253\ $MeV, the first order region does not expand
substantially while in the range above $\mu _{0}=379\ $MeV, it expands
beyond the critical mass at $\mu =0$. Also, the results of Figs. \ref%
{2D_eta_gD} and \ref{2D_35gD} suggest that the range of $\mu _{0}$ that 
corresponds to the curvature from lattice QCD is insensitive to the size of $g_{%
\mathrm{D}}^{\ast }\left( 0\right) $ or $m_{\mathrm{c}}(0)$ in our model. 
\textbf{\ }

\begin{figure}[tbh]
\begin{center}
\textbf{%
\includegraphics[height=3.0in,keepaspectratio,angle=0]{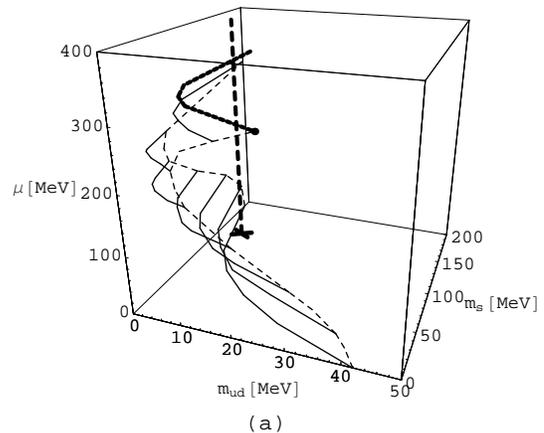} %
\includegraphics[height=3.0in,keepaspectratio,angle=0]{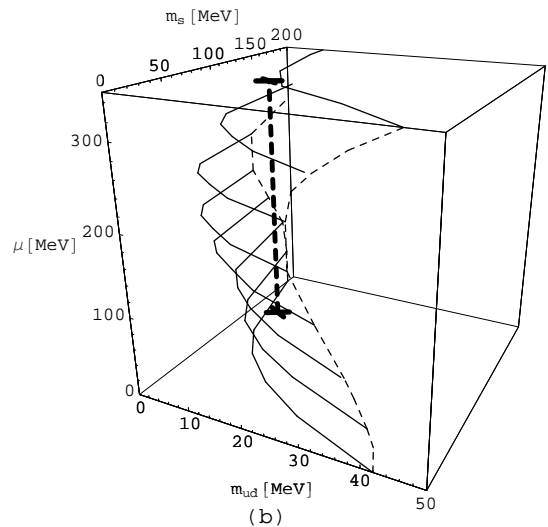} }
\end{center}
\caption{The critical surface in the $(\protect\mu ,m_{\mathrm{ud}%
},m_{\mathrm{s}})$ space with the 't Hooft coupling $g_{\mathrm{D}}^{\ast
}\left( 0\right) =3.5g_{\mathrm{D}}$ to better describe the critical line in
lattice at $\protect\mu =0$. The panel (a) is for $\protect\mu %
_{0}=253$ MeV and the panel (b) is for $\protect\mu _{0}=379$ MeV. The
physical quark mass lines are denoted by the thick dashed lines. The
physical quark mass line intersects the critical surface which means the CEP
exists in the panel (b). In contrast, CEP does not exist in the panel (a). 
In the panel (a), the critical surface has a boundary curve above
which we have no phase transition. It looks like a parallel curve but
actually it has a small but finite dependence on $\mu$.
}
\label{3D_35}
\end{figure}

To further explore the existence of the CEP within this model, one needs to
go beyond the $m_{\mathrm{ud}}=m_{\mathrm{s}}$ limit. \textbf{\ }Again,
using $g_{\mathrm{D}}^{\ast }\left( 0\right) =3.5g_{\mathrm{D}}$, we show
the critical surface in the $(\mu ,m_{\mathrm{ud}},m_{\mathrm{s}})$ space
with $\mu _{0}$ constrained by lattice data. In the panel (a) of
Fig. \ref{3D_35}, $\mu _{0}=253\ $MeV, the $U_{\mathrm{A}}(1)$ restoration
is strong enough such that the critical surface does not intersect the
physical quark mass line which is denoted by the thick dash line in the
plot. Thus, CEP does not exist in this case. In the panel (b) ($\mu
_{0}=379\ $MeV), on the other hand, the critical surface intersects the
physical quark mass line. Thus, the CEP exists. \textbf{\ }

A few comments are in order:

(a) The most interesting feature of our result is the back-bending behavior
of the critical surface. In our model we use $\mu _{0}$ to parameterize the $%
U_{\mathrm{A}}(1)$ restoration and fit it to lattice data. But similar
behavior was also seen in another model with the repulsive vector-vector
current interaction added to the NJL model \cite{Fukushima:2008is}. Thus,
the back-bending seems to be a generic feature of the NJL model, and it
could even be a property of QCD. This raises the possibility that despite
the shrinking of the first order region at lower $\mu $, the QCD critical end
point might still exist due to the expansion at higher $\mu $. In this case,
it might be challenging for lattice QCD calculations using 
the analytic continuation and the Taylor expansion to probe the back-bending
behavior. 
It will be worthwhile to investigate how 
precise the lattice computation should be in order to detect or rule out the
back-bending scenario.

(b) As mentioned above, the reason for shrinking of the first order region is 
uncertain yet. Both partial $U_{\mathrm{A}}(1)$ restoration and the
vector-vector repulsive four-fermion interaction can give this effect. 
Lattice computations could, however, test whether the $U_{\mathrm{A}}(1)$
restoration is responsible for the shrinking by computing $m_{\eta ^{\prime
}}$ or the topological susceptibility at small $\mu $.

\section{\label{sec:Summary}Summary}

We have discussed the chiral phase transition in hot and dense QCD with
three light flavors. Inspired by the well-known fact that the $U_{\mathrm{A}%
}(1)$\ anomaly could induce first order phase transitions, we have studied
the effect of the possible restoration of the $U_{\mathrm{A}}(1)$ symmetry
at finite density. In particular, we explored the link between the $U_{%
\mathrm{A}}(1)$ restoration and the recent lattice QCD results of de
Forcrand and Philipsen, in which the first order phase transition region
near zero chemical potential ($\mu $) shrinks in the quark mass and $\mu $
space when $\mu $ is increased. Starting from the Ginzburg-Landau theory for
general discussions, we then used the Nambu--Jona-Lasinio model for
quantitative studies. With the partial $U_{\mathrm{A}}(1)$\ restoration
modeled by the density dependent 't Hooft interaction, we have fit the
shrinking of the first order region found in de Forcrand and Philipsen's
lattice calculation at low $\mu $. At higher $\mu $, the first order region
might shrink or expand, depending on the scenarios. This raises the
possibility that despite the shrinking of the first order region at lower $\mu 
$, the QCD critical end point might still exist due to the expansion at
higher $\mu $. In this case, very high precision lattice data will be needed
to detect the back-bending of the critical surface with the currently
available analytic continuation or Taylor expansion approaches.
Finally, since the $\eta'$ mass and the topological susceptibility are
sensitive to the strength of the $U_{\mathrm{A}}(1)$ anomaly, lattice
computations of these quantities at small $\mu$ could check whether
the strength of the $U_{\mathrm{A}}(1)$ anomaly is reduced when $\mu$
is increased. These calculations, however, have to be carried out using
either imaginary chemical potential or derivative expansions due to
the fermion sign problem with finite $\mu$.

\begin{acknowledgments}
The authors are grateful to Naoki Yamamoto for fruitful comments.
JWC thanks YITP at the Kyoto University and the INT at the University of
Washington for hospitality. JWC, KO and UR are supported by the NSC and NCTS
of Taiwan. KF is supported, in part, by Japanese MEXT grant No.\ 20740134
and the Yukawa International Program for Quark Hadron Sciences.
\end{acknowledgments}

\end{document}